\documentclass[12pt,preprint]{aastex}

\def\ltsima{$\; \buildrel < \over \sim \;$}
\def\gtsima{$\; \buildrel > \over \sim \;$}
\def\lsim{\lower.5ex\hbox{\ltsima}}
\def\gsim{\lower.5ex\hbox{\gtsima}}
\def\lapp{\ifmmode\stackrel{<}{_{\sim}}\else$\stackrel{<}{_{\sim}}$\fi}
\def\gapp{\ifmmode\stackrel{>}{_{\sim}}\else$\stackrel{<}{_{\sim}}$\fi}

\newdimen\minuswidth    
\setbox0=\hbox{$-$}
\minuswidth=\wd0
\catcode`@=\active
\def@{\kern\minuswidth}
\setbox0=\hbox{\rm0}
 
\shorttitle{} 
\shortauthors{Dalessandro et al.}
 
\begin{document} 
\title{Multiple populations in the old and massive Small Magellanic Cloud globular cluster NGC~121
   \footnote{Based on observations collected with
    the NASA/ESA {\it HST}, obtained at the Space Telescope Science
    Institute, which is operated by AURA, Inc., under NASA contract
    NAS5-26555 and collected at the ESO-VLT under the program 086.D-0665.}  }

\author{
E. Dalessandro\altaffilmark{2,3},
E. Lapenna\altaffilmark{2,3},
A. Mucciarelli\altaffilmark{2,3},
L. Origlia\altaffilmark{3},
F. R. Ferraro\altaffilmark{2},
B. Lanzoni\altaffilmark{2}
}
\affil{\altaffilmark{2} Dipartimento di Fisica \& Astronomia, Universit\`a degli Studi
di Bologna, viale Berti Pichat 6/2, I--40127 Bologna, Italy}

\affil{\altaffilmark{3} INAF --- Osservatorio Astronomico di Bologna, via Ranzani 1, I-40127, Bologna, Italy}

%

\begin{abstract}
We used a combination of optical and near-UV Hubble Space Telescope photometry and 
FLAMES/ESO-VLT high-resolution spectroscopy
to characterize the stellar content of the old and massive globular cluster (GC) NGC~121 in the Small Magellanic Cloud (SMC).
We report on the detection of multiple stellar populations,  
the first case in the SMC stellar cluster system. 
This result enforces the emerging scenario in which the presence of multiple stellar populations 
is a distinctive-feature of old and massive GCs regardless of the environment, as far as the light element 
distribution is concerned. 
We find that second population (SG) stars are more centrally concentrated than first (FG) ones.
More interestingly, at odds with what typically observed in Galactic GCs, 
we find that NGC~121 is the only cluster so far to be dominated by FG stars that account 
for more than $65\%$ of the total cluster mass.
In the framework where GCs were born with a $90-95\%$ of FG stars,
this observational finding would suggest that either NGC~121 experienced a milder stellar mass-loss with
respect to Galactic GCs or it formed a smaller fraction of SG stars.  
\end{abstract}

\keywords{Magellanic Clouds -- globular clusters: individual (NGC~121) -- stars: abundances --
  stars: RGB -- techniques: photometric, spectroscopic}


\section{Introduction}

All relatively massive ($>4-5 \times 10^4 M_{\odot}$) and old ($10-13$ Gyr) Galactic globular clusters (GCs) studied so far
host multiple stellar populations 
showing appreciable differences in the abundance of He and several other light elements 
\citep[e.g., C, N, Na, O, Al, Mg, see][for a review]{gratton12}. 
Hereafter, we define these multiple populations as light element MPs (LE-MPs), to distinguish them from the 
multiple populations showing large iron abundance differences, 
as those observed in $\omega$Cen \citep[see e.g.][]{pancino00,ferraro04,johnson10} 
and in Terzan~5 \citep{ferraro09,massari12,origlia13}.
These LE-MPs manifest themselves 
as multi-modal or broadened evolutionary sequences in color-magnitude-diagrams (CMDs), 
when appropriate filters (or filter combinations) are adopted 
\citep[see, for example,][]{piotto07,marino09,monelli13,dalex11a,dalex14a,milone13,milone15}.

Different scenarios have been proposed over the years to explain the formation of LE-MPs. 
In the most popular formation models, LE-MPs are the result of 
a self-enrichment process, which likely occurred in the very early epochs of GC evolution ($\sim 100$ Myr). 
A second generation (SG) formed from a combination of 
the ejecta of stars from a first population (polluters - FG) and from a ``pristine material'' 
\citep{decressin07,dercole08,demink09,conroy12,denis14}.
However, all of the proposed scenarios face serious problems as none can explain more than few relevant observations 
\citep[e.g.][]{salaris14,bastian15,renzini15}.

As a matter of fact we still lack a self-consistent explanation 
of the physical process(es) at the basis of the multiple population formation and evolution. 
To this end it is necessary to use a thorough and wide approach that combines photometric, spectroscopic 
and kinematical information.  
For example, crucial insights on the very first epoch of GC evolution can be obtained by observing and
comparing clusters over a wide range of properties (metallicities, mass, structural parameters etc.) and 
in different environments (from dwarf to giant elliptical galaxies).
One of the most debated and still poorly understood topics is what determines the fraction of SG stars in GCs. 

By using the available collection of spectroscopic and photometric data, \citet{bastian15} showed that 
the fraction of SG stars is remarkably uniform ($N_{\rm SG}/N_{\rm TOT}\sim0.68\pm0.07$) regardless of the GC mass, 
metallicity, and present-day galactocentric distance, with the only exceptions of 
NGC~6362 \citep{dalex14a,mucciarelli16} and NGC5272 \citep{massari16} that show equally-populated FG and SG. 
To reproduce this evidence, models require that all GCs had an initial mass up to 10-100 times larger than 
the current one,
and they eventually 
lost $>90-95\%$ of their original mass through tidal-stripping or 
gas-expulsion \citep{dercole08,conroy12}. This strong requirement, which is common to any model, 
is typically known as the 
``mass-budget problem''  and it may have important implications on the galaxy mass-assembly scenarios.
However, it should be also noted that it appears to be inconsistent with some observational results (e.g. \citealt{larsen12}, 
\citealt{bastian15}) and GC dynamical evolution models (e.g.  \citealt{krui15}).

On the same line, important efforts have been made to detect LE-MPs in other galaxies. The presence of LE-MPs has been directly observed
by means of photometry and/or spectroscopy of resolved stars in the old GCs of the Large Magellanic Clouds 
\citep[see, for example,][]{mucciarelli09} and in the Fornax dwarf galaxy 
\citep{larsen14}, while it has been indirectly inferred in the massive GCs of M31 \citep{schiavon13} and M87 \citep{chung11}
thanks to integrated spectro-photometric studies.
Quite surprisingly, similar investigations still lack for the relatively close ($d\sim60$kpc) 
Small Magellanic Cloud (SMC) irregular dwarf galaxy.

The SMC harbors a population of GCs with an almost continuous distribution in age up to $t\sim 8$ Gyr 
\citep{harris04,carrera08,dias10}. The only significant exception is the older cluster NGC~121, 
a system located at $\sim$ 2.3$^{\circ}$ NW
from the optical center of the SMC \citep{crowl01}. It has a mass $M\simeq 3\times 10^{5} M_{\odot}$ \citep{mackey03}
and age $t_{\rm AGE}\sim 11$ Gyr \citep{glatt08}. 
These properties make 
NGC~121 an extremely interesting target, as {\it it is the only old, Galactic-like GC in the SMC}.

In this work we present a spectro-photometric analysis of NGC~121 with the aim of characterizing its light element
chemical patterns, the presence of LE-MPs (as traced by C, N, O, N, Mg and Al abundances) and their properties. 
The paper is structured as follows: in Section~2 the observational 
data-base is presented, in Sections~3 and 4 the results obtained by means of the photometric and spectroscopic data-sets, respectively,
are discussed. In Section~5 we summarize the most relevant results and draw our conclusions.


\section{Observations and data reduction}

In order to investigate the possible presence of LE-MPs in NGC~121, 
we analyzed HST high-resolution photometric data and 
some high-resolution UVES-FLAMES spectra.

\subsection{Analysis of the photometric data-set}

The photometric data-base used in this work consists of a combination of HST images obtained with
the Wide Field Camera 3/Ultraviolet and Visible Channel (WFC3/UVIS).

The WFC3/UVIS data-set (Prop: 13435, PI: Monelli) consists of 10 images, four exposures with $t_{\rm exp}=200$ s 
acquired through the $F438W$ band, four with $t_{\rm exp}=1061$ s acquired through the $F336W$ and two images in the 
$F814W$ band with $t_{\rm exp}=100$ s. The cluster core is centered in chip$\#2$ in a sub-sample of five images ($2\times F438W$, 
$2\times F336W$ and $1\times F814W$), and in chip$\#1$ in the remaining five. The two sub-groups of images are
rotated by $\sim 30^{\circ}$.

The images have been processed, flat-field corrected and bias-subtracted by using standard HST pipelines 
(\texttt{\textunderscore flt} images). Pixel-area effects have been accounted for by applying the most updated pixel-area-maps 
(\texttt{PAM} images) to each image by means of \texttt{IRAF} tasks\footnote{IRAF is 
distributed by the National Optical Astronomy Observatory, which is operated by
the Association of Universities for Research in Astronomy, Inc., under cooperative agreement with the National Science Foundation}.

The photometric analysis has been performed independently on each chip by using \texttt{DAOPHOTIV} \citep{alls}.
For each image we selected several tens of bright and relatively isolated 
stars to model the point spread function (PSF). We used a Moffat analytic function and a second-order spatial variation for the PSF.
A first star list has been obtained for each image by independently fitting all the star-like sources, as detected by using a 
relatively shallow threshold at $6\sigma$ from the local background. 
We then used \texttt{ALLFRAME} \citep{allf} on all the WFC3/UVIS images, starting from an input list of stars detected in at least 
four WFC3/UVIS images.

The final star-lists obtained for each image and chip have been then crossed-matched by using \texttt{DAOMATCH}. 
For each filter, single magnitude estimates were homogenized and their weighted mean and standard deviation were finally calculated
by means of \texttt{DAOMASTER}. The final catalog consists of stars measured in the three bands.

To independently assess the quality and completeness of the obtained photometry in the innermost and densest regions of NGC~121,
we complemented the WFC3/UVIS data
with archival Advanced Camera for Survery/High Resolution Channel (ACS/HRC) data. 
The ACS/HRC set (Prop: 10369, PI: Gallagher) consists of 12 images, six obtained in the $F555W$ band, four 
with $t_{\rm exp}=300$ s and two
with $t_{\rm exp}=25$ s, and six $F814W$ images, four with $t_{\rm exp}=260$ sec and two short-exposures
with $t_{\rm exp}=20$ s. These images homogeneously cover the innermost $\sim 20\arcsec$ of the cluster.
To exploit the improved PSF
sampling provided by the ACS/HRC data, in the region where the ACS/HRC and WFC3/UVIS images overlap,
we forced the positions of stars identified in at least five ACS/HRC images to be fitted in the WFC3/UVIS ones 
\citep[see, for example,][]{dalex14b}. We verified that the results obtained with both approaches are comparable, 
therefore for clarity and sake of homogeneity in the following we will use only the WFC3/UVIS catalog.  
 
Instrumental magnitudes have been converted to the VEGAMAG photometric
system by using the prescriptions and zero-points reported on the dedicated HST web-pages
\footnote{http://www.stsci.edu/hst/wfc3/analysis/uvis\textunderscore zpts}. 
The instrumental positions of stars have been
separately roto-translated to the absolute ($\alpha$, $\delta$) coordinates 
by using as secondary astrometric standards, calibrated onto the 2MASS system, the stars in common with our own 
near infrared SOFI/NTT photometric catalog 
(this is part of the same survey presented in \citealt{mucciarelli09b}; we refer to that paper for details about 
the photometric analysis).

\subsection{Analysis of the spectroscopic data-set}

Five giant stars  located at a radial distance of $20\arcsec<r<40\arcsec$
from the cluster center \citep{glatt08} 
have been observed with UVES-FLAMES \citep{pasquini02} under the program 086.D-0665 (PI: Mucciarelli)
using the grating 580 Red Arm CD\#3,
which provides a high spectral resolution (R$\sim$40000) and a spectral coverage
between 4800 and 6800$\rm\mathring{A}$.
The target stars have been selected from our own
near-infrared SOFI/NTT photometric catalog
in the brighter portion of the RGB ($K_{s}\sim$13-14),
and considering only targets without close companions with comparable or brighter magnitude.

The spectra have been acquired in a series of 13 exposures of $\sim$ 45 min each.
The data reduction was performed by using the dedicated ESO pipeline
including bias subtraction, flat fielding, wavelength calibration,
spectral extraction and order merging.
Single spectra of each target have been finally co-added, reaching a S/N per
pixel of $\sim$30 at $\sim$6000$\rm\mathring{A}$.
Identification number, right ascension, declination, $J, H, K_{s}$ magnitudes and distance
from the cluster centre of the observed stars are listed in Table~\ref{tab1}.

The radial velocity of the targets has been determined with \texttt{DAOSPEC} \citep{stetson08} 
through the measure of about 200 metallic lines per star. 
The average radial velocity obtained for NGC~121 is $v_{r}$ = +144.8 km s$^{-1}$ ($\sigma$ = 2.0 km s$^{-1}$), 
in good agreement with previous determinations. 
In particular, from the analysis of integrated light spectra of NGC~121 \citet{zinn84} found 
       $v_{r}$ = +139 $\pm$ 20 km s$^{-1}$,
      \citet{hesser86} $v_{r}$ = +138 $\pm$ 15 km s$^{-1}$ and \citet{dubath97} 
      $v_{r}$ = +147 $\pm$ 2 km s$^{-1}$ while
      \citet{dacosta98} found $v_{r}$ = +138 $\pm$ 4 km s$^{-1}$ by using CaII triplet.
Based on the measured radial velocities, 
the five stars are cluster members.

Effective temperatures (T$_{eff}$) and surface gravities (log~$g$) have been calculated by using 
the photometric information.
T$_{eff}$ have been derived from the ($J-K$)$_{0}$-T$_{eff}$ calibration of \citet{alonso99},
assuming $E(B-V)$ = 0.03 \citep{schlegel98,schlafly11} and extinction coefficients from \citep{mccall04} for the 2MASS photometric system.
Surface gravities have been obtained with the Stefan-Boltzmann
equation by adopting ($m-M$)$_{0}$ = 19.02 mag \citep{glatt08}. Bolometric corrections
were computed following the prescriptions by \cite{buzzoni10} and adopting an average stellar mass of 0.8$M_{\odot}$, obtained
from a BaSTI isochrone \citep{pietrinferni04} with an age of 11 Gyr and a metallicity Z=0.002,
in agreement with the metallicity estimated in this work (see Section~4). 
Finally, the microturbulent velocities have been derived by requiring 
no trend between FeI abundances and the line strengths.

The chemical abundances of Fe, Na and Mg have been calculated from the measured 
equivalent widths (EW) by using
the package \texttt{GALA}\footnote{http://www.cosmic-lab.eu/gala/gala.php} \citep{mucciarelli13a}. 
The measurements of EWs
have been obtained through \texttt{DAOSPEC}, iteratively launched by
the \texttt{4DAO}\footnote{http://www.cosmic-lab.eu/4dao/4dao.php} software \citep{mucciarelli13b}.
We have build a specific linelist for the target stars by using suitable synthetic spectra computed
at the UVES resolution and using as guess parameters the photometric estimates.
The lines have been carefully selected in order to avoid
possible blends and adopting the atomic data of the last release of the
Kurucz/Castelli compilation\footnote{http://www.user.oats.inaf.it/castelli/linelists.html}.
Suitable model atmospheres have been obtained by running the 
\texttt{ATLAS9}\footnote{http://www.user.oats.inaf.it/castelli/sources/atlas9codes.html} 
code and then used as input in the computation of 
the synthetic spectra with the \texttt{SYNTHE} package \citep{sbordone05}.

The Fe abundances have been obtained measuring $\sim$100 Fe~I lines, 
the abundances of Na have been obtained from the lines at 5682, 5688 and 5889$\rm\mathring{A}$
and those of Mg~I from the transition at 5711$\rm\mathring{A}$.
In order to take into account non-local thermodynamical equilibrium 
effects the Na abundances have been corrected according to \cite{gratton99}.

We used the spectral synthesis technique, employing the package \texttt{SALVADOR} (Mucciarelli et al., in prep), 
which follows the method already described in \citet{mucciarelli12} to derive the abundances of O (because the forbidden oxygen line at 6300.3 $\rm\mathring{A}$ 
is contaminated by a close Ni line)
and to estimate upper limits for the abundances of Al (because 
the Al lines at 6696-6698 $\rm\mathring{A}$ are too weak to be properly measured).
We have checked that the [OI] line is free from telluric contamination and
we adopted a proper C and N abundances according to the estimate of \cite{gratton00}
for RGB stars close to the tip.
Finally, we adopted the solar reference values of \citet{grevesse98} for Na, Al, Mg and Fe
and of \citet{caffau11} for O. 
Table~\ref{tab2} lists the atmospheric parameters and the derived iron abundances, while Table~\ref{tab3} lists 
some abundance ratios of interest.

For the computation of internal uncertainties two main sources of error have been considered:

(1) the error arising from the EW measurements has been computed by dividing the
line-to-line dispersion by the square root of the number of lines used.
Thanks to the high number of available lines for FeI we obtained internal
uncertainties lower than 0.02 dex.
For the Na abundances we obtained uncertainties in the range 0.03-0.06 dex.
For Mg abundances, based on only one line, we assumed as abundance uncertainties 
those obtained from the EW errors provided by DAOSPEC.
For O abundances, obtained through spectral synthesis, we resorted to Monte Carlo simulations 
following the method described by \citet{mucciarelli13c}.

(2) the error arising from atmospheric parameters has been estimated
by taking into account the photometric errors, which yield
an average uncertainty for T$_{eff}$ of $\sim$ 100 K. This value translates in
an uncertainty on log~$g$ of $\sim$ 0.05 dex.
According to these values, we finally estimated that a reasonable uncertainty
for the microturbulence is 0.1 km~s$^{-1}$.
The typical final uncertainties have been obtained by varying each parameter independently
and the cumulative one was computed by summing in quadrature the different terms.
We found $\sigma_{\rm Fe} = 0.09$ dex, $\sigma_{\rm O} = 0.03$ dex, $\sigma_{\rm Na} = 0.1$ dex,
and $\sigma_{\rm Mg} = 0.08$ dex. 

\section{Results from the photometric data-set}
\label{phot}

\begin{figure}[h]
\centering
\includegraphics[angle=0,scale=0.8]{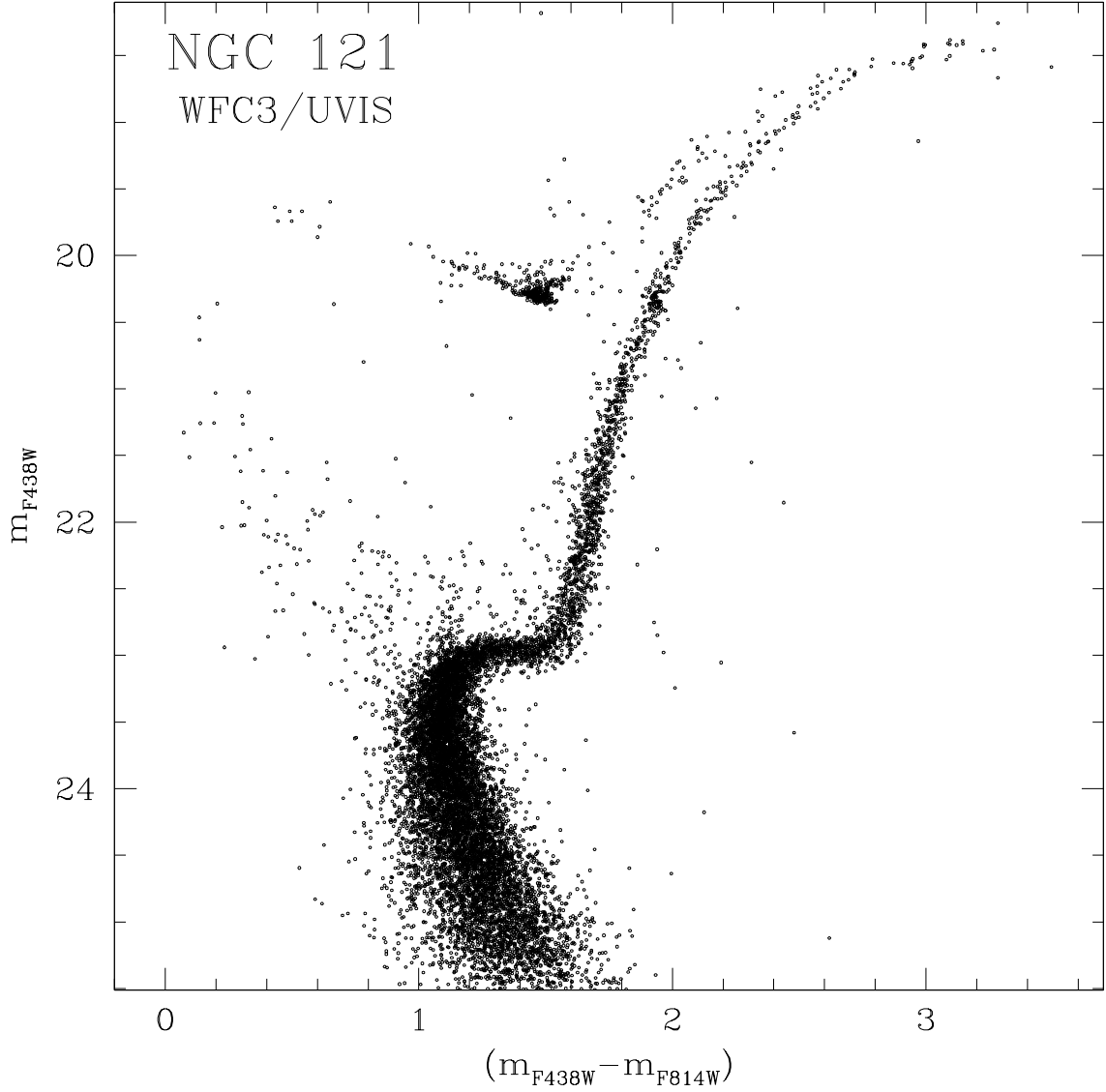}
\caption{($m_{\rm F438W}, m_{\rm F438W}- m_{\rm F814W}$) CMD of NGC~121 as obtained from the HST WFC3/UVIS data-set.}
\label{cmd1}
\end{figure}

The ($m_{\rm F438W}, m_{\rm F438W}-m_{\rm F814W}$) CMD of NGC121 obtained with the WFC3/UVIS data is shown in Figure~\ref{cmd1}.
All sequences are well defined and they appear to be properly sampled from the RGB-tip to $\sim2$ magnitudes 
below the MS turnoff. As expected, given both the position of NGC~121 in the SMC and the fact 
that we are sampling only the innermost regions 
of the cluster, very little contamination from field stars appears in the CMD as a sparse sequence 
of young MS stars overlapping the cluster's Blue Straggler star population.

To search for the presence of LE-MPs we analyzed the ($m_{\rm F336W}$, $m_{\rm F336W}$ - $m_{\rm F438W})$ CMD, 
which is well known to be sensitive to variations of the NH and CN-CH molecular bands \citep{sbordone11}. 
Indeed, in the CMD of Figure~\ref{cmd2}, the color broadening of the RGB significantly 
exceeds the one expected purely from photometric errors, and a hint of sequence split appears quite clearly 
at $m_{\rm F336W}<22$.
It has been demonstrated 
\citep{marino08,sbordone11,dalex14a,mucciarelli16}
that in the ($m_{\rm F336W}$, $m_{\rm F336W}$ - $m_{\rm F438W})$ CMD stars enhanced in N and Na and depleted in O 
and C  (SG) populate the red part of the RGB, while
the Na/N-poor and O/C-rich stars (FG) occupy the blue part.

To maximize the separation between different sub-populations, we used the HST version of the color index 
$C_{\rm U,B,I}$ \citep{monelli13}, defined as ($C_{F336W, F438W, F814W}$) = $(m_{\rm F336W}-m_{\rm F438W})-(m_{\rm F438W}-m_{\rm F814W})$. 
This color combination is sensitive to both light metals and He variations \citep{sbordone11,monelli13} 
and it has been proven to effectively separate LE-MPs in GCs. The ($m_{\rm F336W}$, $C_{F336W, F438W, F814W}$) CMD is shown in the left panel of
Figure~\ref{color}. In this diagram, for $m_{\rm F336W}< 22$ the RGB clearly splits in two sequences, with
the redder sequence corresponding to O-rich/Na-poor (FG), while the bluer sequence being associated to
the O-poor/Na-rich (SG) sub-populations \citep{monelli13}.

In order to better distinguish the sub-populations,   
for RGB stars in the magnitude range $20.45 <m_{\rm F336W}< 22.00 $ we computed the distribution of 
the color distance from the
RGB mean ridge line\footnote{To derive the RGB mean ridge line we considered only as sub-sample of 
``fiducial'' RGB stars defined as those 
falling in the hand-drawn box in the left panel of Figure~\ref{color}. We then computed the median color of RGB stars 
 in bins of 0.3 magnitudes and interpolated them with a spline.} ($\Delta C_{{F336W,F438W,F814W}}$, hereafter $\Delta C$) in the ($m_{\rm F336W}$, $C_{F336W, F438W, F814W}$) CMD 
(Figure~\ref{color}, right panels). 
The distribution is clearly bimodal. To test the statistical significance of
the bimodality, we performed the Gaussian Mixture Modeling (GMM) test \citep{muratov10} finding that, as
expected, the hypothesis of unimodality can be rejected with a probability $>99.9\%$.
The distribution is well fitted by two Gaussians with peaks at $\Delta C=-0.055\pm0.004$ mag and
$\Delta C=+0.037\pm0.005$ mag and both with a dispersion $\sigma\sim0.025$,
which is compatible with the color errors in the considered magnitude range.
We classified all stars with values of $\Delta C$ larger or smaller than 0 as FG and SG sub-populations
respectively.
With this selection we count 457 FG and 247 SG stars in the WFC3/UVIS field of view, corresponding to the $65\pm4\%$
and $35\pm3\%$ of the total. At odds with what typically found so far in Galactic GCs, NGC~121 is dominated by FG stars. 
It is important to note also that, as the HST field of view ensures a sampling of $\sim70\%$ of the entire cluster 
extension
($r_t\sim165\arcsec$, \citealt{glatt09}), we can consider these population ratios as a good approximation of the global ones.  
  
We estimated the RGB width ($W_{RGB}$), which is defined as the extension of the 
$\Delta C$ distribution, after removing the 5 per cent bluest and reddest stars.
We found $W_{RGB}\sim0.16$ mag, which turns out to be 
slightly larger than what previously obtained for clusters with a similar metallicity (see the case of NGC~288 which has 
$W_{RGB}=0.14$ mag; \citealt{monelli13}). 
Given the almost linear correlation between 
the [Na/Fe] and $\Delta C$ distributions found by \citet{monelli13}, this observational finding 
would suggest a wider [Na/Fe] dispersion in NGC~121 than in Galactic GCs with similar metallicity 
where this analysis has been performed.

\begin{figure}[h]
\centering
\includegraphics[angle=0,scale=0.8]{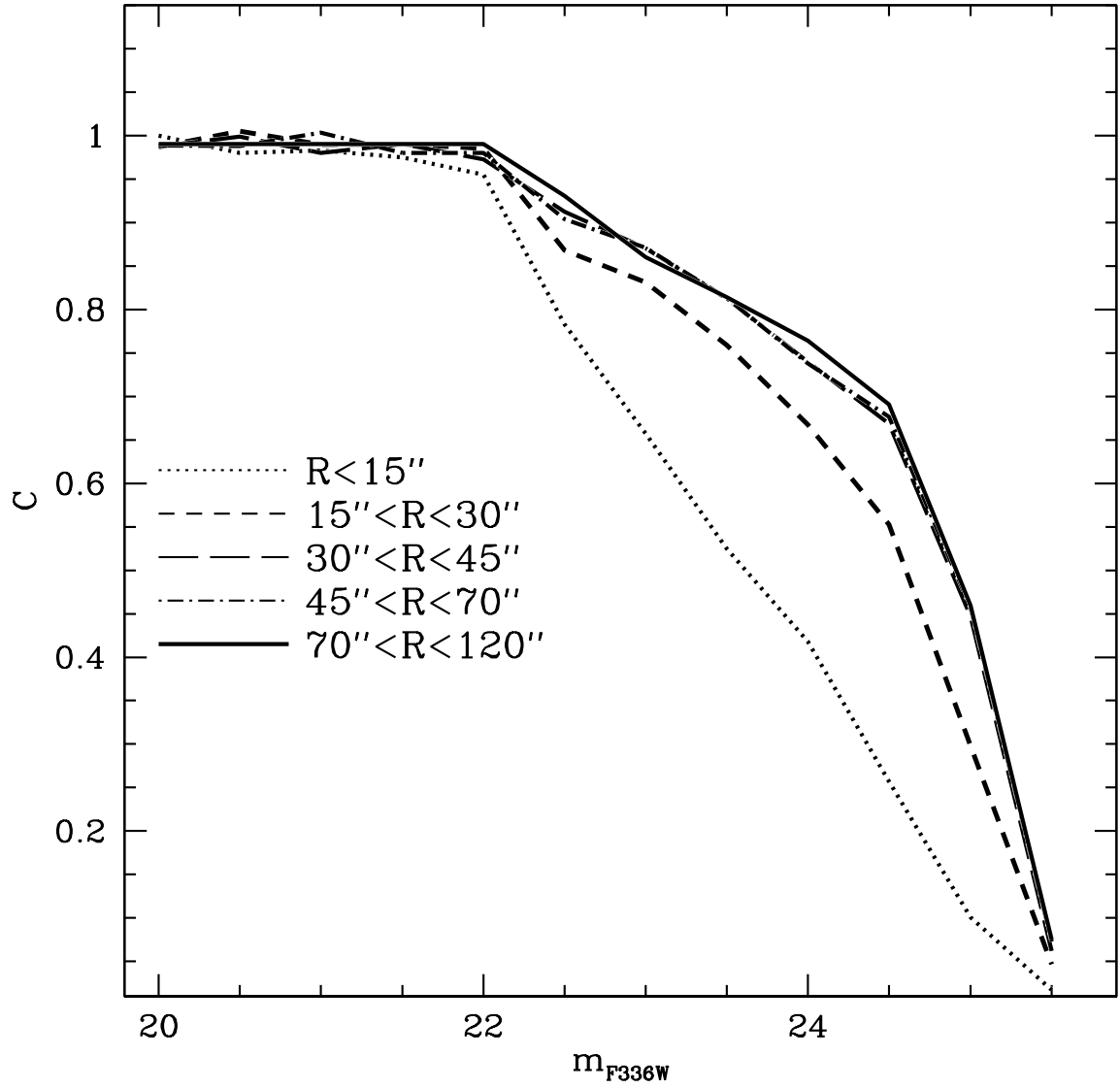}
\caption{Photometric completeness C as a function of the $m_{\rm F336W}$ magnitude for different radial bins (see labels).
}
\label{comple}
\end{figure}

To asses the completeness of our catalog we performed extensive artificial star experiments following the
approach described in Dalessandro et al. (2011b; 2015). Briefly, we first derived mean ridge lines in the 
($m_{\rm F336W}$, $m_{\rm F336W}$ - $m_{\rm F438W})$ and ($m_{\rm F336W}$, $m_{\rm F336W}$ - $m_{\rm F814W})$ CMD, 
adopting 0.5 mag-wide bins in $m_{\rm F336W}$ and
selecting the corresponding median values obtained after a 2$\sigma$-clipping rejection. Then we generated a catalog of
simulated stars with a $m_{\rm F336W}$-input magnitude extracted from a luminosity function modeled to reproduce the
observed one. To each star extracted we assigned input magnitudes  $m_{\rm F438W}^{in}$ and $m_{\rm F814W}^{in}$ 
by means of an interpolation along the mean ridge line. 
To prevent any artificial increase of the crowding conditions, we simulated only one artificial star within a $25\times25$ pixel 
box at each run. This procedure was performed by using the same PSF models and reduction procedures described in Section~2.1.
The artificial star catalog thus obtained was used to derive the completeness curves as the ratio $C=N_{\rm out}/N_{\rm in}$
between the number of stars recovered after the photometric reduction ($N_{\rm out}$) and the number of simulated stars 
($N_{\rm in}$) in each magnitude bin. The completeness curves as a function of the $m_{\rm F336W}$ magnitude at different distances
from the cluster center are shown in Figure~\ref{comple}. In the range of magnitude where LE-MPs have been identified
the completeness $C\sim100\%$. 
 
The photometric data-set allows to sample the cluster area 
within $\sim 120\arcsec$, corresponding to $\sim4.5$ half-light radii ($r_h=27.1\arcsec$, \citealt{glatt08}). 
The cumulative radial distribution of the FG and SG sub-populations is shown in Fig.~\ref{rad}.
It clearly reveals that SG stars are significantly
more centrally concentrated than the FG sub-population, as assessed by means of a Kolmogorov-Smirnov test.
This is in agreement with what observed in most Galactic GCs studied so far, apart 
from NGC~6362 \citep{dalex14a}, NGC~6752 and NGC~6121 \citep{nardiello15}, where FG and SG 
are totally spatially mixed, and M~15 \citep{larsen15}, where FG have been suggested to be more segregated than SG stars. 

The radial distribution of the ratio $N_{\rm SG}/N_{\rm TOT}$ between SG and the total number of stars 
(Figure~\ref{rad}, bottom panel) 
indicates that SG stars are more centrally concentrated, showing a maximum  
$N_{\rm SG}/N_{\rm TOT}\sim 0.45$ at $r<10\arcsec$ and progressively decreasing values moving outward,
down to a minimum value, $N_{\rm SG}/N_{\rm TOT}\sim 0.10$ at $r\sim 60\arcsec$.

\section{Results from the spectroscopic data-set}
\label{spec}

The chemical analysis of NGC~121 provides an average iron abundance 
[Fe/H] = $-1.28$ dex ($\sigma$ = 0.06 dex), with no hint of intrinsic iron spread.
This value is higher than previous derivations 
by \citet{mighell98} who found [Fe/H] = $-1.71$ dex, \citet{dacosta98} [Fe/H] = $-1.51$ dex,
\citet{zinn84} [Fe/H] = $-1.46$ dex, and \citet{johnson04} [Fe/H] = $-1.41$. However, these analysis were based
on photometry or low-resolution spectra that are significantly more prone to uncertainties and to systematics related to the 
adopted calibrations.

No intrinsic spreads have been found also for light-element abundances [O/Fe], [Na/Fe] and [Mg/Fe], as confirmed 
using the Maximum Likelihood algorithm described in \citet{mucciarelli12} that rules out 
any significant light-element spread in the sample.

Figure~\ref{anti} shows the location of the five RGB stars measured in NGC~121 (blue circles)
in the [O/Fe]-[Na/Fe] and [Mg/Fe]-[Al/Fe] diagrams, 
in comparison with the individual stars measured in the Galactic GCs 
\citep[gray points, data from][]{carretta09_gir,carretta09_uves} 
and in the old LMC GCs \citep[red squares,][]{mucciarelli09}.
At variance with the stars in the Galactic and LMC old GCs, that 
share the same mean loci
and cover a large range in the light-element abundances, 
the stars of NGC~121 cover a small area in the two planes 
shown in Figure~\ref{anti}, pointing out that these stars are indeed internally homogeneous.
At a first order approximation, the position of these stars in both the [O/Fe]-[Na/Fe] and [Mg/Fe]-[Al/Fe] diagrams 
is consistent with that of Galactic and LMC old globular FG stars.
We note however that they do not follow exactly the distribution of MW and LMC GCs, especially in the [Mg/Fe]-[Al/Fe] diagram. 
This might be related to the different chemical evolution history of the SMC. 

This finding suggests that the observed spectroscopic sample is composed only by FG stars.
Although the analyzed sample is admittedly small, 
clear evidence of light-element spread and some anti-correlations were found by \citet{mucciarelli09}
in old LMC clusters, based on a similar number of stars (5-7).
However the lack of SG stars in the spectroscopic sample of NGC~121 is qualitatively compatible with the 
fact this is dominated by FG stars that account for $\sim70\%$ of the total population at the surveyed distances.


\section{Discussion and Conclusions}

\begin{figure}[h]
\centering
\includegraphics[angle=0,scale=0.8]{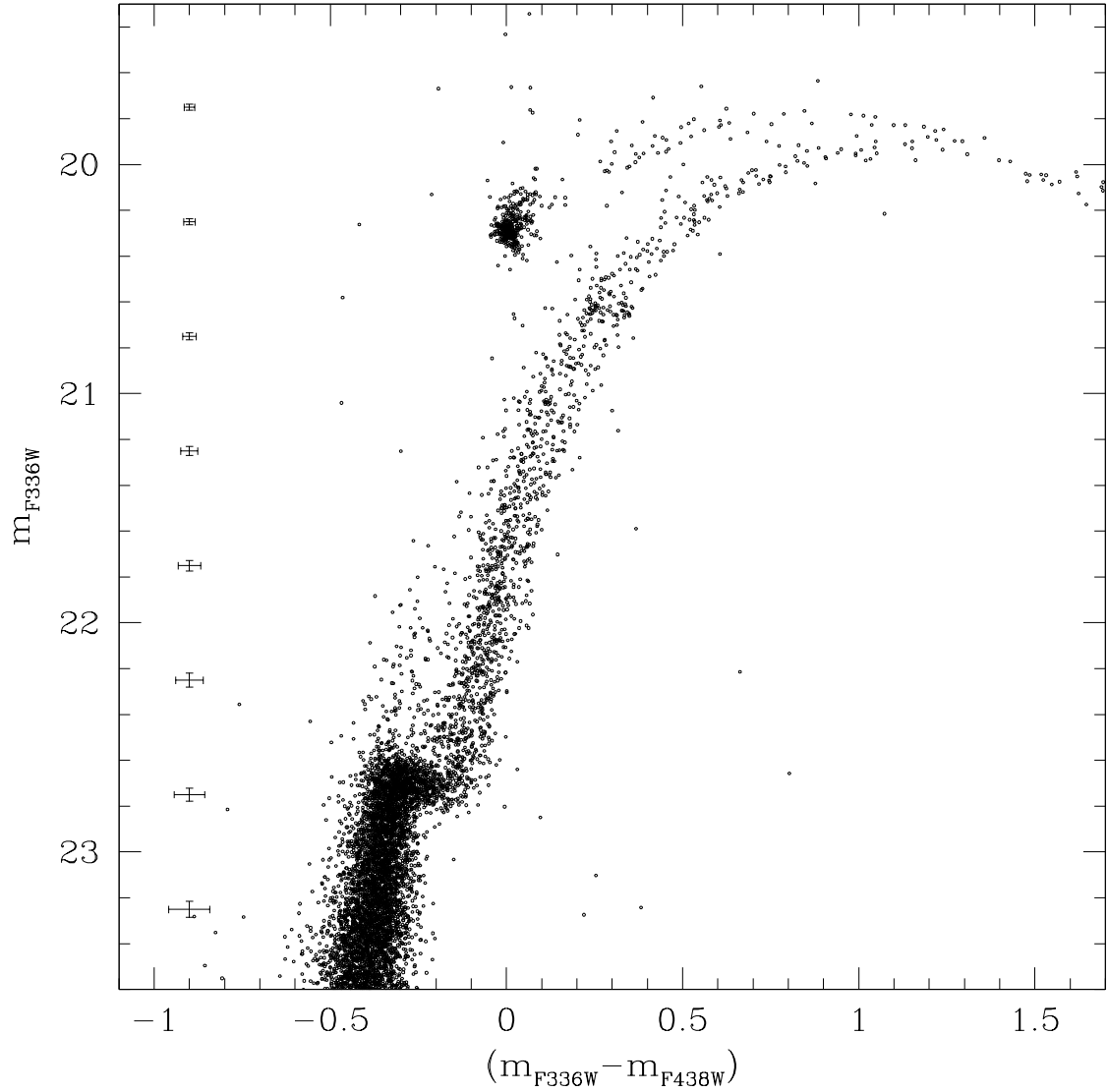}
\caption{($m_{\rm F336W}, m_{\rm F336W}- m_{\rm F438W}$) CMD of NGC~121 as obtained from the HST WFC3/UVIS data-set. 
The RGB sequence is broader than expected from photometric errors and for 
$m_{\rm F336W}<22$ a hint of splitting is apparent. In this CMD, Na-rich and O-poor SG stars would be 
located on the red sequence, while
Na-poor and O-rich (FG) stars lie on the blue one. }
\label{cmd2}
\end{figure}

\begin{figure}[h]
\centering
\includegraphics[angle=0,scale=0.8]{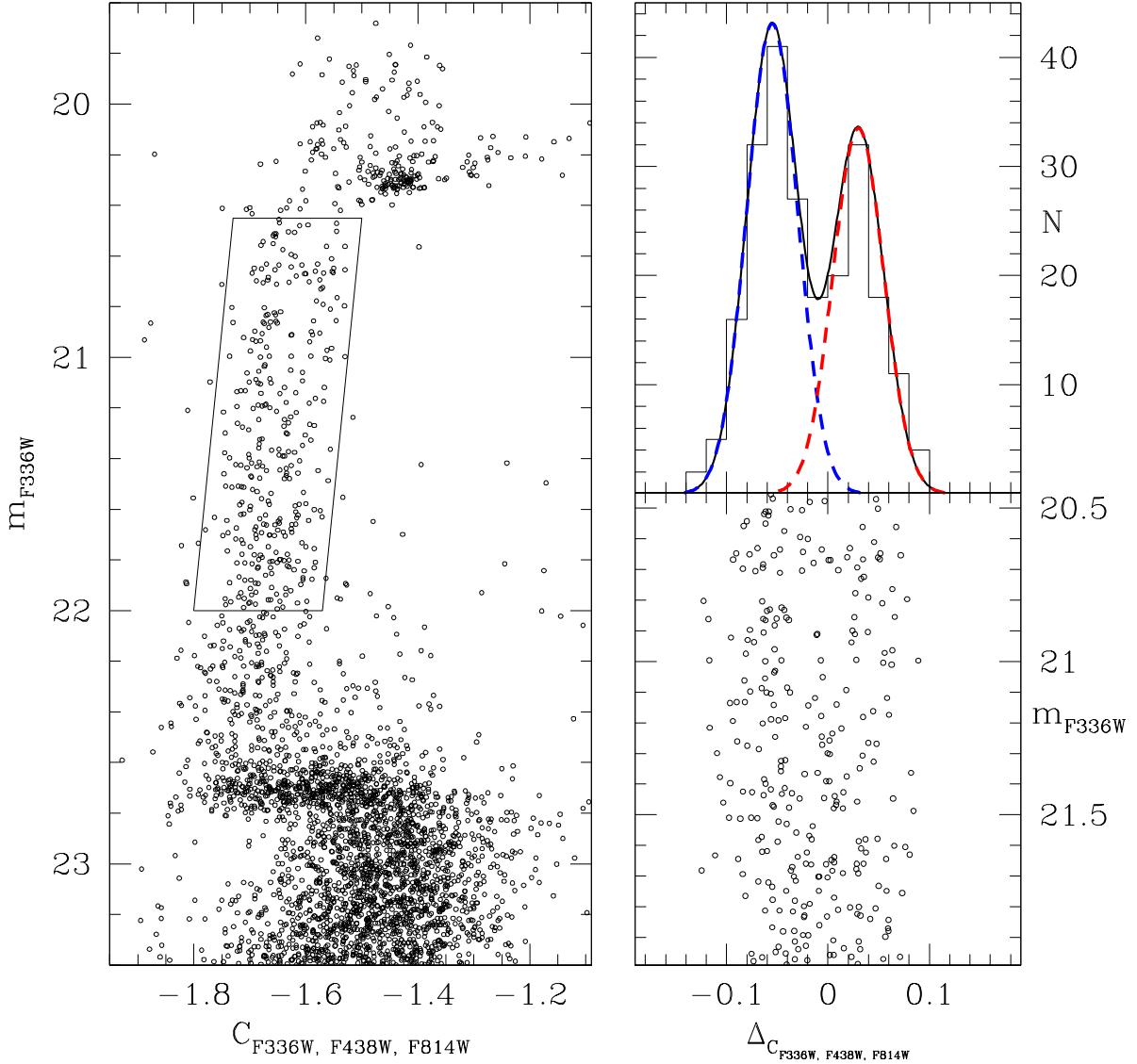}
\caption{
{\it Left panel}: ($m_{\rm F336W}, C_{\rm F336W,F438W,F814W}$) CMD. The box marks the region 
where the RGB splitting is more evident and stars have been selected to compute the differential color distribution 
($\Delta_{\rm C_{\rm F336W, F438W, F814W}}$).
{\it Right lower panel}: distribution of the color distance ($\Delta_{\rm C_{\rm F336W, F438W, F814W}}$) 
of RGB stars in the magnitude interval $20.45<m_{\rm F336W}<22.00$ with respect to the RGB ridge mean line. 
{\it Right upper panel}: histogram of the color distribution.
}
\label{color}
\end{figure}

\begin{figure}[h]
\centering
\includegraphics[angle= 0,scale=0.8]{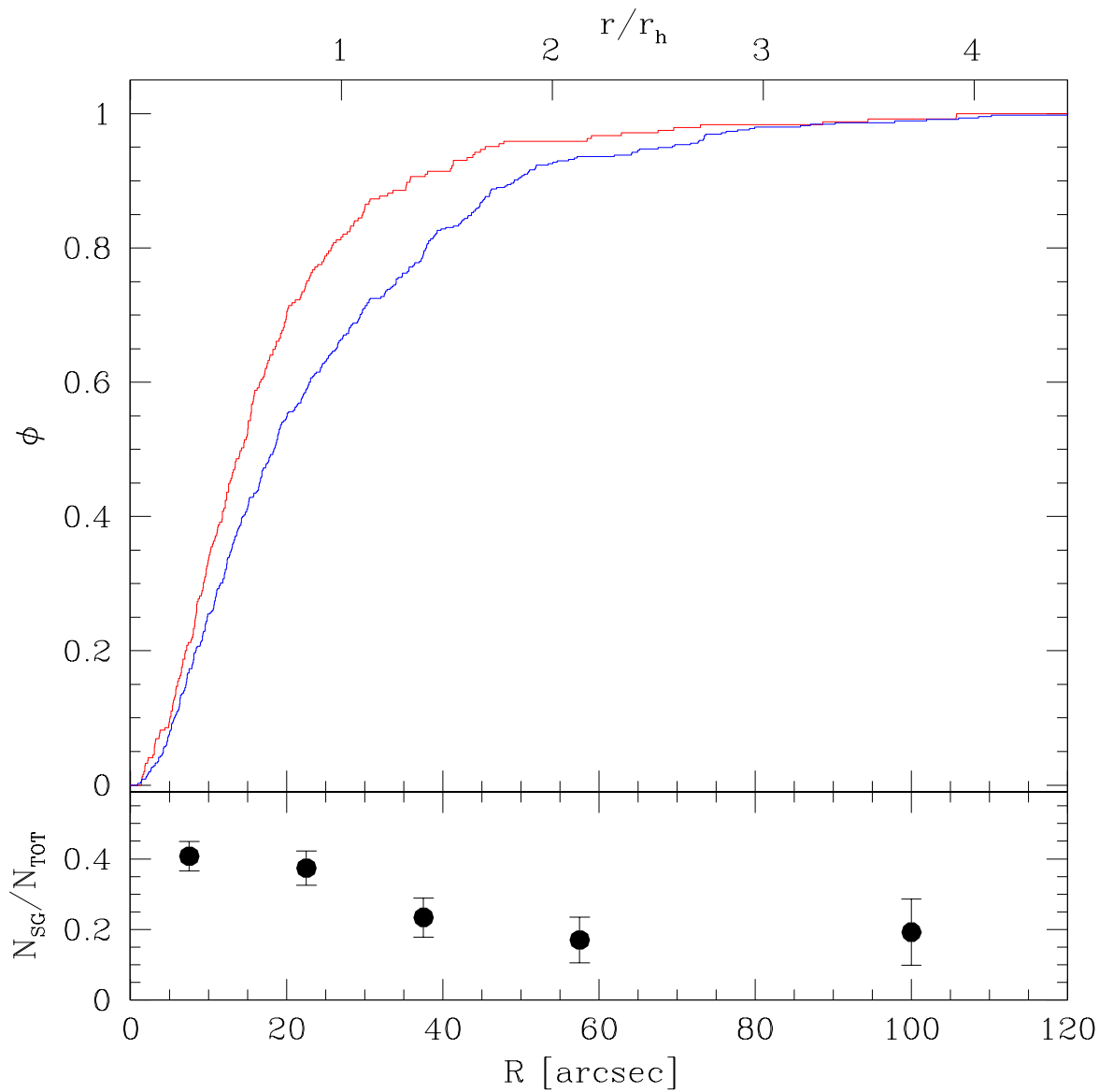}
\caption{{\it Upper panel}: Cumulative radial distribution of FG (blue) and SG (red) stars within the WFC3 field of view selected as
detailed in Section~\ref{phot} as a function of the distance from the cluster center in arcsec and in units of $r_h$. 
{\it Lower panel}: $N_{\rm SG}/N_{\rm TOT}$ as a function of the distance from the cluster center.}
\label{rad}
\end{figure}

\begin{figure}[]
\centering
\includegraphics[angle=0,scale=0.4]{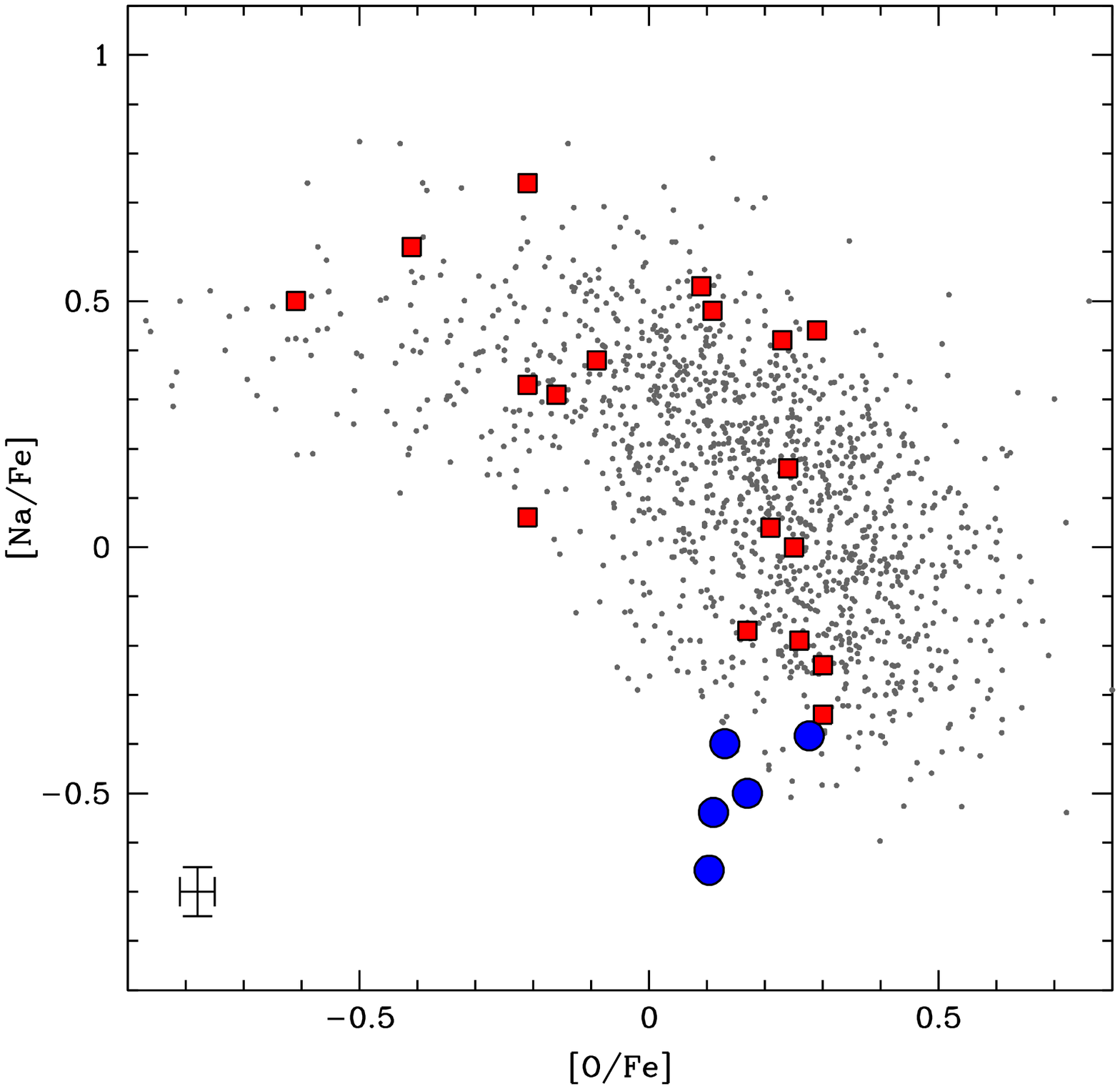}
\\
\includegraphics[angle=0,scale=0.4]{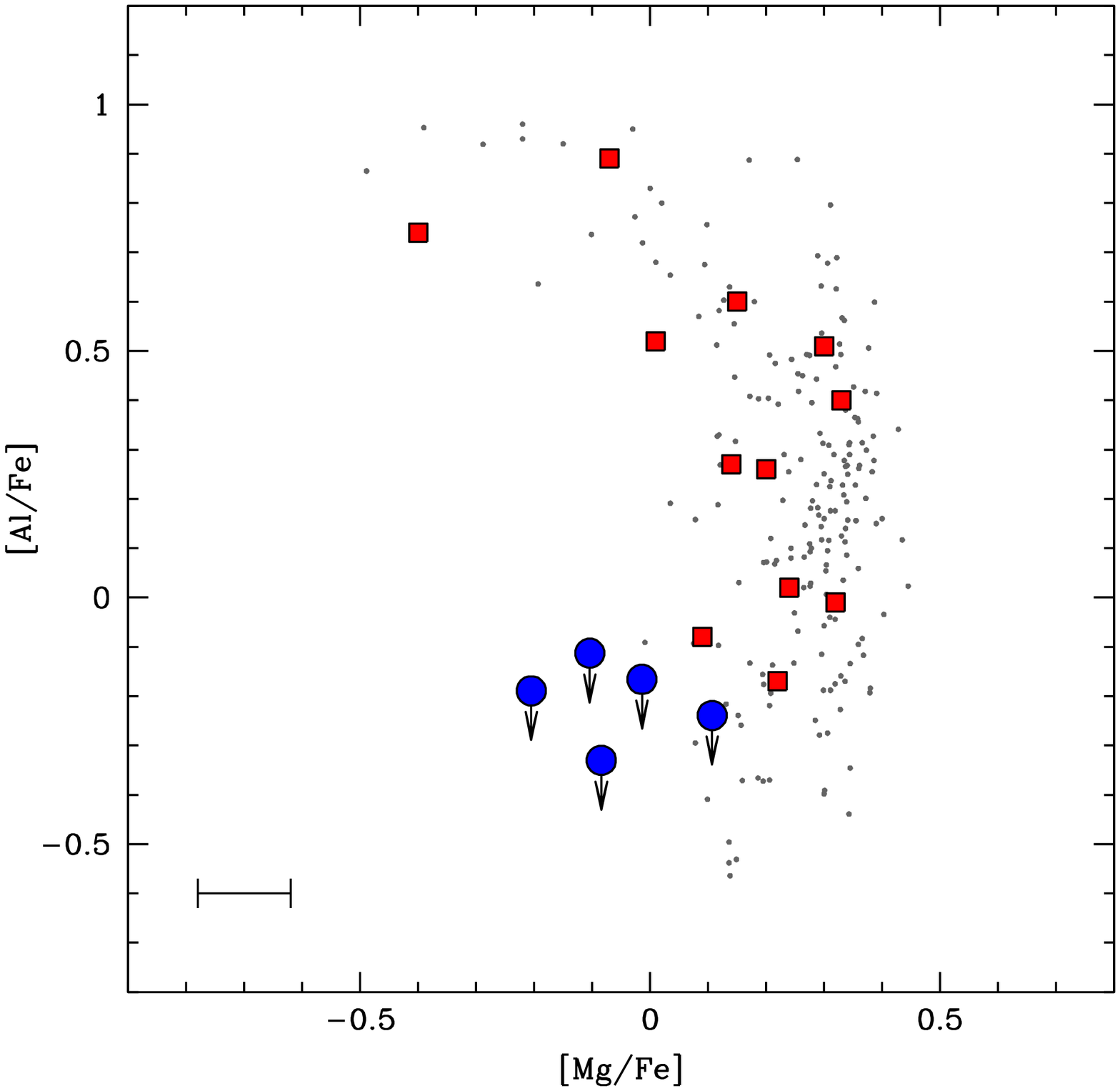}
\caption{[O/Fe]-[Na/Fe] (top panel) and
[Mg/Fe]-[Al/Fe] (bottom panel) diagrams.
The blue solid circles are the target stars in NGC~121. For Al only upper limits have been derived.
Red symbols mark stars in old LMC GCs from \citet{mucciarelli09}.
Gray dots mark the values for stars in Galactic GCs \citep{carretta09_gir,carretta09_uves}.
}
\label{anti}
\end{figure}

The spectro-photometric analysis of the old and massive SMC cluster NGC~121 performed in this paper, 
provided the following main results:

\begin{itemize}
\item{NGC~121 shows a clear broadening and/or splitting of the RGB sequence in CMDs where the F336W 
filter is used (Section~\ref{phot}; Figures~\ref{cmd2} and \ref{color}).  
This result represents the first evidence that NGC~121 hosts LE-MPs. 
This cluster is also the first system to show such a property in the SMC, 
thus enforcing the emerging scenario in which LE-MPs are a distinctive-feature of old and massive GCs, 
regardless of the environment. In fact, this feature is observed in the Milky Way, in the Fornax dwarf galaxy and, now, in both the Magellanic Clouds.}

\item{SG stars are more centrally concentrated than FG stars, as typically observed in Galactic GCs (Section~\ref{phot};
Figure~\ref{rad}).}

\item{NGC~121 is dominated by FG stars that account for $65\%$ of the entire population 
of the cluster, at variance with what found in Galactic GCs, where the fraction of FG stars is 
smaller than the SG one.} 

\item{The analysis of high-resolution spectra of five RGB stars yields an average iron abundance 
[Fe/H] = $-1.28$ dex ($\sigma$ = 0.06 dex), with no hints of intrinsic iron spread (Section~\ref{spec}). 
Homogeneous light-element abundance ratios ([O/Fe], [Na/Fe] and [Mg/Fe] --
Figure~\ref{anti}) have been also found, consistent with all the analyzed stars being O-rich/Na-poor FG stars.}

\end{itemize}

\citet{dercole08} and \citet{vespe13} suggested that at birth GCs should be populated by $\sim95\%$ of FG stars.
Then they reach present-day number ratios because of a preferential FG star loss due to both early 
stellar ($t<1-2 $ Gyr) and long-term dynamical evolutions. 

\begin{figure}[h]
\centering
\includegraphics[angle= 0,scale=0.8]{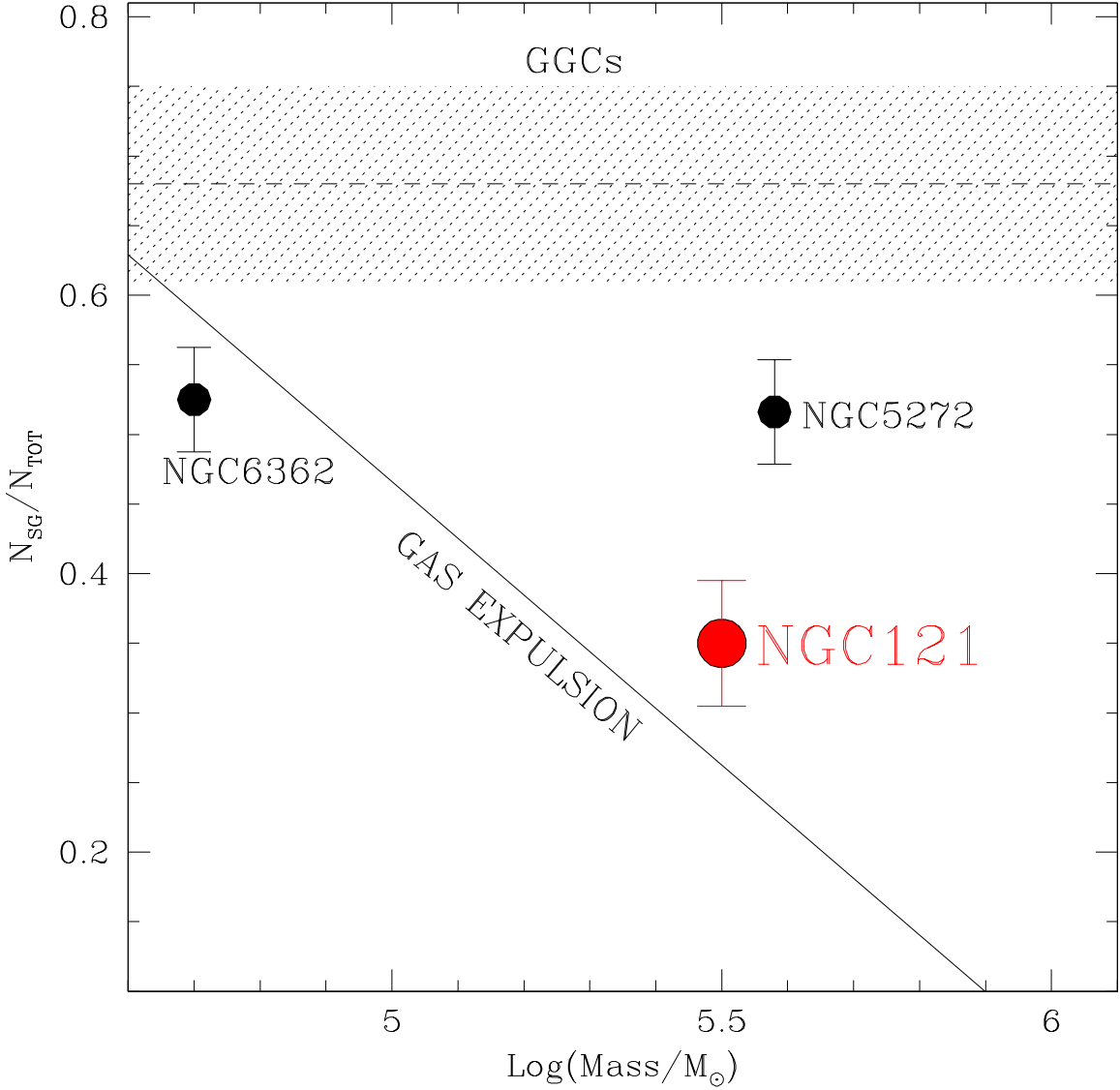}
\caption{Fraction of SG stars as a function of GC mass. 
The values for NGC~121 (this work) and for 
NGC~6362 (Dalessandro et al. 2014a) and NGC~5272 (Massari et al. 2016) inferred from photometric 
indicators are marked as big dots.
The horizontal dashed line  
marks the number ratio mean value and the dashed area the $1\sigma$ region for Galactic GCs, as estimated by
\citet{bastian15} using a collection of spectroscopic and photometric values available in the literature. 
The solid line is the prediction by \cite{khalaj15} for gas expulsion driven mass-loss.}
\label{mass}
\end{figure}

Figure~\ref{mass} shows the location of NGC~121 
in the Mass {\it vs} $N_{\rm SG}/N_{\rm TOT}$ diagram, 
together with the values measured in 
NGC~6362 \citep{dalex14a,mucciarelli16} and NGC~5272 \citep{massari16},
by using the same photometric technique to distinguish among different sub-populations.
For comparison we also show the average 
$N_{\rm SG}/N_{\rm TOT}$ ratio computed by  Bastian \& Lardo (2015)
using a collection of spectroscopic and photometric 
measurements for 33 Galactic GCs, which stays constant at a value of $(68\pm7\%)$.  

It is worth noticing that the SG fraction observed in NGC~121 appears to be consistent
with the value predicted by the gas-expulsion model of \cite{khalaj15} based on N-body simulations (Figure~\ref{mass}), 
which foresees a decreasing  SG star fraction with increasing the cluster mass. Such a similarity is quite remarkable, 
as at the mass of NGC~121, these models underpredict by a factor of three or more the fraction of SG stars in Galactic GCs
(Figure~\ref{mass}). 

A dominant FG sub-population in NGC~121 can be explained with 
a less severe mass loss (still $\sim90\%$) experienced by this cluster, when compared to Galactic GCs,
due to weak tidal interactions with the intrinsically shallower potential of the SMC, and to the fact that 
this GC orbits at a quite large distance 
(about 2.5 kpc) from the SMC center.
Another possible explanation is that NGC~121 lost the same fraction of FG stars as Galactic GCs, but 
formed a smaller fraction of SG stars, 
because of, e.g., a less efficient self-enrichment processes, gas retention, the impact of different environments on 
LE-MP formation, etc.
 
However, on top of these speculative arguments, it should be noted that the available FG and SG population ratios 
for Galactic GCs can be quite uncertain and 
not well representative of the entire cluster population. 
In fact, spectroscopic-based number ratios strongly depend on the
the ability to separate different sub-populations that sometimes have not discretely different abundance patterns.
Moreover, they are spatially 
incomplete by construction, typically not sampling the cluster central regions.
On the other hand, photometric derivations are generally limited to the small HST field of view and cannot account 
for possible large-scale radial variations.

Such potential biases in the derived population ratios in most of the Galactic GCs 
are a serious concern. Indeed, robust estimates obtained in
NGC~6362 (Figure~\ref{mass}; \citealt{dalex14a}) and NGC~5272 \citep{massari16},
for which global number ratios are available (thanks to a proper combination of HST and ground-based
wide-field observations), 
indicate $N_{\rm SG}/N_{\rm TOT}\sim50\%$, significantly lower 
than the average value of $\sim 0.7$ computed by \citet{bastian15}.

Concluding, NGC~121 turns out to be an interesting and intriguing case among the old, massive GCs studied so far.
Understanding the origin of its peculiarities, will likely provide important insights into the 
formation and early evolution of GCs and multiple populations. Moreover, performing similar studies on a larger sample of 
clusters will allow us to shed new light on the role of the environment and host galaxy properties 
on the multiple population properties.


\acknowledgements

We thank the anonymous referee for his/her comments and suggestions.
L.O. and E.L. acknowledge the PRIN-INAF 2014 CRA 1.05.01.94.11: 
``Probing the internal dynamics of globular clusters. The first
comprehensive radial mapping of individual star kinematics with the 
new generation of multi-object spectrographs'' (PI: L.
Origlia).


{}


\begin{deluxetable}{rrrccccc}
\tablewidth{0pc}
\tablecolumns{8}
\tablecaption{Identification number, right ascension, declination, J, H, K magnitudes, distance from the cluster center 
and radial velocity of the target stars.}
\tablehead{\colhead{ID} & \colhead{R.A.} & \colhead{Decl.} & \colhead{J} & \colhead{H} 
& \colhead{K} & \colhead{Distance} & \colhead{$v_{r}$} \\
& \colhead{(J2000)} & \colhead{(J2000)} & \colhead{(mag)} & \colhead{(mag)} &
\colhead{(mag)} & \colhead{(arcsec)} & \colhead{(km s$^{-1}$)}}
\startdata
 & & & & & & & \\
  9  &  6.6842639  &  --71.5367593  &  13.977  &  13.262  &  13.111  &  21.7  &  143.90 $\pm$ 0.12  \\
 14  &  6.7033858  &  --71.5295107  &  14.253  &  13.560  &  13.440  &  25.8  &  142.17 $\pm$ 0.13  \\
 18  &  6.7233897  &  --71.5450034  &  14.371  &  13.677  &  13.552  &  37.7  &  145.25 $\pm$ 0.16  \\
 31  &  6.6845726  &  --71.5315588  &  14.671  &  14.024  &  13.966  &  28.2  &  147.91 $\pm$ 0.12  \\
 35  &  6.6970187  &  --71.5477942  &  14.716  &  14.048  &  13.934  &  41.0  &  146.23 $\pm$ 0.26  \\
\enddata
\tablecomments{The distances have been computed by adopting the cluster center coordinates from \cite{glatt08}.}
\label{tab1}
\end{deluxetable}

\begin{deluxetable}{rccccc}
\tablewidth{0pc}
\tablecolumns{6}
\tablecaption{Atmospheric parameters and [FeI/H] abundances for the target stars.}
\tablehead{\colhead{ID} & \colhead{$T_{eff}^{phot}$} & \colhead{log~$g^{phot}$} & \colhead{$v_t^{spec}$} &
\colhead{[FeI/H]} \\
& \colhead{(K)} & \colhead{(dex)} & \colhead{(km s$^{-1}$)} & \colhead{(dex)} & }
\startdata
  9  &  3874  &  0.36  &  1.30  &  --1.20 $\pm$ 0.09    \\
 14  &  3904  &  0.48  &  1.45  &  --1.36 $\pm$ 0.09    \\
 18  &  4010  &  0.57  &  1.35  &  --1.28 $\pm$ 0.09    \\
 31  &  4105  &  0.70  &  1.25  &  --1.31 $\pm$ 0.09    \\
 35  &  4159  &  0.77  &  1.25  &  --1.24 $\pm$ 0.09    \\
\enddata
\tablecomments{Identification number, photometric temperatures and gravities,
microturbulent velocities, and [FeI/H] abundance ratios with internal uncertainty.
For all the stars a global metallicity of [M/H]= $-1.25$ dex has been assumed for the model
atmospheres. The reference solar value is taken from \cite{grevesse98}.}
\label{tab2}
\end{deluxetable}

\begin{deluxetable}{rcccrccccccc}
\tablewidth{0pc}
\tablecolumns{12}
\tablecaption{Chemical abundances for light- and $\alpha$-elements of the target stars.}
\tablehead{\colhead{ID} & \colhead{[O/Fe]} & \colhead{[Na/Fe]} &  \colhead{[Mg/Fe]} & 
\colhead{[Al/Fe]}  \\
& \colhead{(dex)} & \colhead{(dex)} & \colhead{(dex)} & \colhead{(dex)} & }
\startdata
  9  &  0.10 $\pm$ 0.04  &  --0.66 $\pm$ 0.01    &  --0.01 $\pm$ 0.08  &  $<$ --0.17   \\
 14  &  0.28 $\pm$ 0.03  &  --0.38 $\pm$ 0.06    &  --0.10 $\pm$ 0.07  &  $<$ --0.11   \\
 18  &  0.11 $\pm$ 0.05  &  --0.54 $\pm$ 0.05    &  --0.21 $\pm$ 0.08  &  $<$ --0.19   \\
 31  &  0.13 $\pm$ 0.04  &  --0.40 $\pm$ 0.02    &    0.11 $\pm$ 0.07  &  $<$ --0.24   \\
 35  &  0.17 $\pm$ 0.05  &  --0.50 $\pm$ 0.05    &  --0.08 $\pm$ 0.08  &  $<$ --0.33   \\
\enddata
\tablecomments{Abundance ratios with internal uncertainty.
The abundances of oxygen and magnesium have been derived from one line,
while for aluminium we estimated an upper limit only.
The reference solar values are taken from \cite{grevesse98} for all species
except for oxygen for which we adopted the value of \cite{caffau11}.}
\label{tab3}
\end{deluxetable}


\end{document}